\documentstyle[12pt]{article}
\begin{document}
\baselineskip=0.7cm
\newcommand{\EQ}{\begin{equation}}
\newcommand{\EN}{\end{equation}}
\newcommand{\EQA}{\begin{eqnarray}}
\newcommand{\EQN}{\end{eqnarray}}
\newcommand{\e}{{\rm e}}
\newcommand{\Sp}{{\rm Sp}}
\renewcommand{\theequation}{\arabic{section}.\arabic{equation}}
\newcommand{\Tr}{{\rm Tr}}
\renewcommand{\thesection}{\arabic{section}.}
\renewcommand{\thesubsection}{\arabic{section}.\arabic{subsection}}
\makeatletter
\def\section{\@startsection{section}{1}{\z@}{-3.5ex plus -1ex minus 
 -.2ex}{2.3ex plus .2ex}{\large}} 
\def\subsection{\@startsection{subsection}{2}{\z@}{-3.25ex plus -1ex minus 
 -.2ex}{1.5ex plus .2ex}{\normalsize\it}}
\makeatother
\def\thefootnote{\fnsymbol{footnote}}
\begin{flushright}
UT-KOMABA/98-10 \\
BROWN-HEP-1122\\
May 1998
\end{flushright}

\vspace{1cm}
\begin{center}
\Large
 Space-Time Uncertainty Principle and Conformal Symmetry \\
 in D-Particle Dynamics

\vspace{1cm}
\normalsize
{\sc Antal Jevicki}
\footnote{
e-mail address:\ \ antal@het.brown.edu}
\\
\vspace{0.3cm}
 {\it Department of Physics, Brown University, Providence}
\\
\vspace{0.5cm}
{\sc Tamiaki Yoneya}
\footnote{
e-mail address:\ \ tam@hep1.c.u-tokyo.ac.jp}
\\
\vspace{0.3cm}
 {\it Institute of Physics, University of Tokyo, Komaba, Tokyo}

\vspace{1.3cm}
Abstract\\

\end{center}
Motivated by the space-time uncertainty principle, 
we establish a conformal symmetry in 
the dynamics of D-particles.  The conformal symmetry,  
combined with the supersymmetric non-renormalization 
theorem, 
uniquely determines the classical form of the 
effective action for a probe D-particle 
 in the background of a heavy D-particle source,  previously 
constructed by Becker-Becker-Polchinski-Tseytlin. 
Our results strengthen 
the conjecture proposed by Maldacena 
on the correspondence, 
 in the case of D-particles,  between the 
supergravity and the supersymmetric Yang-Mills 
matrix models in the large $N$-limit, the latter being 
the boundary conformal field theory of the former in the classical 
D-particle background in the near horizon limit.

\newpage
\section{Introduction}
The dynamics of D-branes at least in the 
low-energy regime is  
described  by the supersymmetric 
Yang-Mills theories.  The 
interpretation of the Yang-Mills theories 
 is entirely different 
from the usual applications to the 
unified theories of particle interactions. 
The Higgs fields which emerge from a part of the 
components of the 
gauge fields as a result of dimensional reduction  
are now identified as describing the transverse coordinates of 
D-branes and  the associated open string degrees of 
freedom. The gauge symmetry is regarded as 
a symmetry which generalizes the statistics symmetry for 
ordinary particles in quantum mechanics.  
Through the correspondence between open and 
closed string theories, the Yang-Mills theories 
so interpreted are supposed to describe even the 
gravity which is necessarily included in the closed string 
theories.  However, it is fair to say that we do not yet have 
appropriate understanding on the fundamental principles 
which explain why such Yang-Mills theories describe the gravity. 

Recently, based on studies \cite{klebanov0} of D-brane interactions, 
a remarkable conjecture \cite{maldacena} has emerged relating 
the Yang-Mills theory in the large $N$-limit and supergravity. 
In particular,  the $N=4$ super symmetric 
Yang-Mills theory in $3+1$ dimensions 
in a strong-coupling regime should be described 
by the type IIB supergravity 
in the anti-de Sitter background 
AdS$_5 \times S^5$corresponding to 
the near horizon geometry of the $p$=3 extremal black hole 
solution.  A more concrete formulation of this 
conjecture has been given in \cite{klebanov}\cite{witten}.  
Basically, the effective Yang-Mills theory of 
D3-branes is identified as the boundary conformal 
field theory of the 4+1 dimensional anti-de Sitter 
space-time. The SO(4,2) symmetry of the 
latter turns into the conformal symmetry of the 
Yang-Mills theory defined at the  3+1 dimensional 
asymptotic boundary of 
the anti-de Sitter space-time.  
The existence of such a boundary field theory is natural, 
since all legitimate observables in general 
relativistic quantum theory must be  defined asymptotically.  
 
On the other hand, one of the most promising 
proposals related to the  
D-brane dynamics is the so-called Matrix theory \cite{bfss}, 
which interprets the 0+1 dimensional 
N=16 super symmetric Yang-Mills theory as 
a non-perturbative formulation of the 
M-theory, assuming the D-particles 
as the fundamental degrees of freedom behind 
the type IIA strings. A justification for such 
a description comes from the assumption of 
the infinite Lorentz boost 
which is expected to ensure the small velocity of D-particles 
along the transverse directions and the 
decoupling of the higher open string modes \cite{seiberg}. 

In the case of Maldacena's conjecture, the 
decoupling of the higher open string modes is 
ensured by taking the $\alpha' \rightarrow 0$ 
limit with the energy $U\equiv {r \over \alpha'}$ 
of the open strings stretched between the D3-branes kept fixed, 
where $r$ is the transverse distance between the 
D3-branes. The large $N$ limit with large fixed $g_s N$ 
is assumed in order to have small curvature $\sim 
{1\over g_{YM}^2N}$ in the string unit with small string 
coupling constant $g_s \, \, (\propto g_{YM}^2) $.   

Then a natural question arises: Is it possible to 
apply the conjecture to D-particle 
quantum mechanics? In view of the  role played by the conformal symmetry in the case of D3-branes, one of the 
crucial questions would then be whether there exists 
any symmetry which takes place of the conformal 
symmetry of the D3 case. The purpose of the 
present paper is to give an affirmative answer to the 
last question.  We will argue that there indeed exists 
in both sides, the type IIA supergravity and 0+1 dimensional 
Yang-Mills matrix model,  some 
extended conformal symmetry.  This suggests that 
the D-particle quantum mechanics may also be 
interpreted as a boundary ``conformal"  field theory 
  in the background of the classical D-particle solution. 

Since at its root our work is strongly motivated by the 
space-time uncertainty principle \cite{ty1, ty2},  which 
has been proposed by one of the present authors 
as a possible principle characterizing the 
short-distance space-time structure of the 
string theory,  let us 
start in the next section by explaining the connection 
between the space-time uncertainty relation and the 
conformal symmetry.  For a recent review of the 
space-time uncertainty principle including 
its application to D-branes, we refer the reader to 
\cite{ty3}.  We can regard the conformal symmetry 
as a mathematical structure characterizing the 
space-time uncertainty principle. 

\newpage
\section{Space-time uncertainty principle and 
conformal symmetries}

The conformal symmetry of the effective Yang-Mills 
theory of the D3-brane comes basically from the 
symmetry under the scale transformation
\EQ
X_i(x_a) \rightarrow X_i'(x_a') =\lambda X_i(x_a) ,
\label{eq21}
\EN
\EQ
x_a \rightarrow x_a' = \lambda^{-1} x_a ,
\label{eq22}
\EN
where $X_i(x_a) \, \, (i=1,\ldots, 6)$ are the 
the Higgs fields representing the 
space-time coordinates and open string fields 
which are transverse to the D3-branes and 
$x_a  \, \, (a=0,1,2,3)$ are the world-volume 
coordinates including the time $x_0=t$. 
The above scalings which are opposite
\footnote{
In the usual field theory language, this is nothing but the 
trivial fact that the dimensionality of the field is negative 
with respect to the length dimension of the base space. 
}
for the transverse and 
longitudinal (including time) directions can be 
regarded as a signature of the space-time uncertainty 
relation 
\EQ
\Delta T \Delta X \sim \alpha' .
\EN
 Here, $\Delta T$ and $\Delta X$ are 
qualitative measures for the effective longitudinal 
(i.e. along the world volume) and transverse 
space-time distances, respectively.  Note that because of the choice of 
the static gauge, the longitudinal distance is directly related to 
the distance in the target space-time. 
The uncertainty relation for D-branes in general  says that 
the long distance phenomena in the (transverse) target space   
is dual to the short distance phenomena in the world volume and 
{\it viceversa}. 
From the view point of the 
space-time uncertainty relation, the transformations 
 (\ref{eq21}) and (\ref{eq22}) 
are the simplest transformation which leaves the 
uncertainty relation invariant, and it is natural to 
interpret the full (super) conformal 
symmetry as constituting  a set of 
more general symmetry transformations 
which keep the uncertainty relation invariant 
and hence characterize  the possible mathematical structure 
behind the uncertainty principle.  

That the conformal 
symmetry and the space-time uncertainty relation 
are related in this way is remarkable,  if one remembers that the 
original proposal of the space-time uncertainty relation 
also came out  as a simple reinterpretation, 
in terms of the space-time language, 
 of the world sheet conformal invariance, or 
more precisely the $s$-$t$ duality which includes the 
open-closed string dualities,  
in perturbative string theories.  
These two conformal symmetries are in a sense dual to each other, 
and may be regarded as 
yet another example of the dual roles of various 
symmetries in the world sheet and target space-time in string theory. 
In contrast to the world sheet conformal symmetry, the 
conformal symmetry for D-branes involves the 
target coordinates explicitly,  
reflecting the fact that the $s$-$t$ duality 
operates for the open fundamental 
strings stretched between the D-branes.  
In the case of fundamental strings, 
the uncertainty relation is valid for the 
effective space-time distances measured along the 
time ($\Delta T$) and spatial ($\Delta X$) directions on the world sheet. 

Once the conformal symmetry is related to the space-time uncertainty 
principle, we should expect that the symmetry is not 
restricted to D3-brane. 
We can indeed consider the transformation of the same nature  
as above for D-particles 
in the form 
\EQ
X_i(t) \rightarrow X_i'(t') =\lambda X_i(t) , \quad t\rightarrow t'=\lambda^{-1}t
\label{eq24}
\EN
where now the index $i$  runs from $i=1$ to $i=9$.  
Note that for D-particles, the ``longitudinal"  distance 
$\Delta T$ literally refers to the time 
 as in ref. \cite{liyo}. 
In this case, the effective super Yang-Mills 
theory is invariant {\it provided}   
we simultaneously  make the transformation of the string coupling constant 
as 
\EQ
g_s \rightarrow g_s'=\lambda^3 g_s .
\label{eq25}
\EN
This scaling property, which is a special case of the 
scaling $g_s \rightarrow g_s'=\lambda^{3-p} g_s $ for general 
D$p$-branes,
 is equivalent to the fact that the 
characteristic spatial and temporal length scales 
of D-particle dynamics  
are fixed by the string coupling as $\Delta X \sim g_s^{1/3} \sqrt{\alpha'}$ \cite{kabat} 
and $\Delta T \sim g_s^{-1/3}\sqrt{\alpha'}$,  respectively.  
Given that the mass of a D-particle is proportional to $1/g_s\sqrt{\alpha'}$, 
 this is a direct consequence of the 
space-time uncertainty relation 
as shown in \cite{liyo}.  Since, as is well known, 
the string coupling constant $g_s$  can  in principle be
treated as a dynamical variable corresponding 
to the vacuum expectation value of the dilaton, 
it is reasonable to regard this property as a symmetry 
in the D-particle dynamics.  Alternatively, we can 
regard the string coupling as a part of the 
background fields, and the symmetry requires to change the 
background and the manifest dynamical degrees of freedom 
simultaneously.  In principle, there should be some mechanism  
which allows us to eliminate the string coupling 
in terms of other genuine dynamical degrees of freedom. 
In the present formulation of the Matrix theory, this is not 
however manifest.   

We also note that the 
D3-brane ($p=3$)  is special in that the string coupling 
is inert under the scale transformation. 
This means that the dynamics of D3-brane involves 
all the scales in both the target and world volume, 
keeping the dual nature of them.  For other ``dilatonic" branes, 
on the other hand, 
the effective scales are fixed by the vacuum expectation values 
of the dilaton.  The space-time 
uncertainty relation itself is valid, however,  irrespective of such 
specialities for general D-branes. 

Motivated by these considerations, we are led to ask ourselves 
whether it is 
possible to generalize the transformations (\ref{eq24}) and (\ref{eq25}) 
into the 0+1 dimensional 
analogue of the full conformal symmetry of the 
3+1 dimensional super Yang-Mills theory.  

The action of the supersymmetric Yang-Mills matrix 
quantum mechanics is 
\EQ
S =\int dt \, \Tr 
\Bigl( {1\over 2g_s\ell_s} D_t X_i D_t X_i + i \theta D_t \theta 
+{1 \over 4g_s\ell_s^5} [X_i, X_j]^2 -
{1\over \ell_s^2}\theta \Gamma_i [\theta, X_i]\Bigr) 
\label{eq26}
\EN
where the covariant derivative is defined by 
$D_t X ={\partial \over \partial t}X +[A, X] $ 
and $\ell_s$ is the string length constant $(\ell_s 
\propto \sqrt{\alpha'})$.
This is obviously invariant under the scale transformations 
(\ref{eq24}) and (\ref{eq25}) provided that the 
gauge field $A$ is transformed as 
$A(t) \rightarrow \lambda A(t') $  and the time translation 
$t\rightarrow t'=t+c, \, \, X_i(t) \rightarrow X_i(t'),  \, \,  A(t) 
\rightarrow A(t') , \, \, 
g_s \rightarrow g_s'=g_s$.   
The corresponding infinitesimal transformations are 
\EQ
\delta_DX_i =  X_i , \, \, \delta_D  A = A ,\, \, \delta_D t =-  t  , \, \, \delta_D g_s =3 g_s 
\label{eq27} ,
\EN 
and 
\EQ
\delta_H X_i =0 , \, \, \delta_H A =0 , \, \, \delta_H t = 1 , \, \, \delta_H g_s =0 ,
\label{eq28}
\EN
respectively. 
 In addition to these trivial 
symmetries, the action is also invariant (up to a 
total derivative as usual) under the 
special conformal transformation given by 
\EQ
\delta_K X_i = 2  t X_i , \, \, \delta_K A= 2  t A , \, \, 
 \delta_K t =-  t^2 , \, \, \delta_K g_s =6  t g_s . 
\label{eq29}
\EN
In all these symmetry transformations, the fermionic 
variable $\theta$ is assumed to be invariant 
(namely, as zero-dimensional scalar with respect to the 
conformal transformation). 
Note that in the above expressions for the 
infinitesimal transformations  we have suppressed 
the infinitesimal parameters and also that time derivatives of 
 $X_i(t)$ do not appear since we have defined the variations 
of the field as $\delta X(t) \equiv X'(t')-X(t)$ to compare the 
transformation rule of the classical solution in  type IIA string 
theory to be discussed in 
the next section. 
The transformations (\ref{eq27}), (\ref{eq28}) and 
(\ref{eq29}) together  form an SU(1,1) algebra. 
\EQ
[\delta_D, \delta_H] = \delta_H , \, \, 
[\delta_D, \delta_K] =-\delta_K , \, \, 
[\delta_H, \delta_K]=2\delta_D  .
\label{eq210}
\EN 
After the special conformal transformation 
the string coupling {\it constant}  
is no more constant.  As remarked before, 
this is acceptable since we regard the string coupling as 
a part of dynamical variables playing the role of a 
background field.   

The operator forms of these generators in the Hamiltonian 
formalism are given as 
\EQ
\tilde{H}
=
{\rm Tr}\Bigl(
{1\over 2}\tilde{\Pi}^2 
-{g_s\over 4}[\tilde{X}_i, \tilde{X}_j]^2 
+\sqrt{g_s}\theta \gamma\cdot [\tilde{X}, \theta]
 \Bigr)+{9\over c}g_s^{7/3}{\partial^2 \over \partial g_s^2} ,
\EN
\EQ
\tilde{K}=-{\rm Tr}\tilde{X}^2  + c g_s^{-1/3} ,
\EN
\EQ
\tilde{D}=-{1\over 4}{\rm Tr}\Bigl(\tilde{X}\tilde{\Pi}
+\tilde{\Pi}\tilde{X} \Bigr)-i 3(g_s{\partial \over \partial g_s})  .
\EN
The field dependent parts of these operators  can easily 
be inferred from the total derivative terms of the 
action under the transformations (\ref{eq27}) 
$\sim$ (\ref{eq29}). 
Here,  c is an arbitrary constant and 
\EQ
(g_s{\partial \over \partial g_s})
\equiv 
{1\over 2}g_s^{7/3}(
{\partial \over \partial g_s}g_s^{-4/3}
+g_s^{-4/3}{\partial \over \partial g_s}) .
\EN
For notational simplicity,  we have set $\ell_s=1$ and 
chosen a particular ordering of $g_s$ and 
${\partial \over \partial g_s}$.  The ordering is however not unique. 
In the present paper, we will not elaborate on this point, since the 
ordering is not important for the following discussions. 

In constructing the closed operator algebra, we adopted a special frame 
for the coordinate fields defined by 
\EQ
X_i = \sqrt{g_s}\tilde{X}_i  ,
\EN
\EQ
\Pi_i= {1\over \sqrt{g_s}}\tilde{\Pi}_i={1\over g_s}D_tX_i
={1\over \sqrt{g_s}}D_t\tilde{X}_i ,
\EN
and  treated the canonical variables with tilde and $g_s$ as independent variables.  
Note that in terms of these variables, 
the space-time scaling transformations (\ref{eq24}) and 
(\ref{eq25}) are 
\EQ
\tilde{X} \rightarrow \lambda^{-1/2}\tilde{X} ,
\EN
\EQ
\tilde{\Pi} \rightarrow \lambda^{+1/2}\tilde{\Pi} ,
\EN
\EQ
g_s\rightarrow \lambda^{3}g_s .
\EN
The algebra is 
\EQ
i[\tilde{D}, \tilde{H}] = \tilde{H} ,
\EN
\EQ
i[\tilde{D}, \tilde{K}] = -\tilde{K} ,
\EN
\EQ
i[\tilde{H}, \tilde{K}]= 2\tilde{D} .
\EN

The transformations of fields induced by 
these generators can be regarded as the 
transformations at $t=0$.  For example, the 
special conformal transformation 
can be expressed as 
\EQ
\delta_K \tilde{X}_i\big|_{t=0} = i[\tilde{K}, \tilde{X}] =0 ,
\EN
\EQ
\delta_K \tilde{\Pi}_i \big|_{t=0} = i[\tilde{K}, \tilde{\Pi}] ,
=2\tilde{\Pi}_i
\EN
\EQ
\delta g_s\big|_{t=0} =i [\tilde{K}, g_s] = 0 ,
\EN
\EQ
\delta p_g\big|_{t=0} =i[\tilde{K}, p_g]=-6g_s ,
\label{ivariationgs}
\EN
where 
\EQ
p_g \equiv 
 i{18\over c}g_s^{7/3}{\partial \over \partial g_s} . 
\EN
These results coincide with the transformations (\ref{eq29}) 
since $\tilde{\Pi}_i\big|_{t=0}=i[\tilde{H}, \tilde{\Pi}_i]={d\over dt}\tilde{X}_i$ and 
$p_g\big|_{t=0}=i[\tilde{H}, g_s]={d \over dt}g_s$.  
 
One of the curious features of the above operators is that 
the time translation generator acquires a kinetic term for the 
string coupling. It is not clear whether this means that 
the time development of the string coupling 
should really be taken into account in the 
dynamics of D-particles.  For our present purposes, 
it is suffice to regard 
these operators just as the infinitesimal generators 
for the conformal transformations which are 
the symmetry of the action (\ref{eq26}) without the kinetic 
term for the string coupling as explained in the 
beginning of the section 2.  From this point of view, 
the time dependence of the transformed string coupling only 
means that the particular time dependence generated by this 
operator can be eliminated by making the conformal 
transformation of the fields $X_i$.  

\vspace{0.5cm}
\section{Conformal symmetry of the classical D-particle 
background in the near horizon limit} 
\setcounter{equation}{0}

As is well known, the classical background solution in the 
type IIA superstring theory 
corresponding to the D-particle can be 
obtained by the dimensional reduction 
from the 11 dimensional plane wave solution \cite{aichel}. 
The solution rewritten in a form which is appropriate for 
10 dimensions \cite{witten2} is 
 \EQ
ds_{11}^2=e^{-2\phi/3}ds_{10}^2  +
e^{4\phi/3}(dx_{11} - A_0 dt)^2
\EN
with  
\EQ
A_0=-{1\over g_s}\bigl( 
{1\over 1+{q\over r^7}}-1\bigr)
\bigr) ,
\EN
where as usual our convention is to use $x_{11}$ for the 
10th spatial coordinates, and the indices $i$ runs only 
through the transverse directions from 1 to 9 ($r=\sqrt{x_i^2}$). 
The dilaton field is given by 
\EQ
e^{\phi}= g_s e^{\tilde{\phi}}
\EN
with 
\EQ
e^{\tilde{\phi}} =\bigl( 1 +{q\over r^7}\bigr)^{3/4}
\EN
and the charge $q$ is given by 
\EQ
q=60\pi^3 (\alpha')^{7/2} g_sN
\EN
for $N$ coincident D-particles.  
Here, we have put $\alpha'$ explicitly. 
The 10 dimensional string frame metric $ds_{10}^2$ is 
\EQ
ds_{10}^2 = -e^{-2\tilde{\phi}/3}dt^2 + e^{2\tilde{\phi}/3}dx_i^2 .
\label{eq36}
\EN

Following \cite{maldacena}\cite{maldacena2}, we now consider the near horizon limit 
$\alpha'\rightarrow 0$, keeping  fixed the 
energy of open strings between the D-particles and the 
probe measuring the metric and also the Yang-Mills coupling 
constant, 
\EQ
U \equiv  {r \over \alpha'} , 
\quad  4\pi^2 g_{YM}^2 \equiv  {g_s\over \alpha'^{3/2}} .
\EN
In this limit, the 10 dimensional metric, the dilaton, 
and the $U(1)$ gauge field $A_{\mu}$ 
which is identified with the Ramond-Ramond 1-form 
gauge field, becomes  
\EQ
ds_{10}^2 =\alpha'\Bigl(-
{U^{7/2}\over \sqrt{Q}}dt^2
+{\sqrt{Q}\over U^{7/2}}
\bigl(dU^2 + U^2 d\Omega_8^2\bigr)\Bigr)  ,
\label{eq38}
\EN
\EQ
e^{\phi} = g_s \Bigl({q \over \alpha'^7 U^7})^{3/4} 
=g_{YM}^2 \Bigl({Q\over U^7}\Bigr)^{3/4} ,
\label{eq39}
\EN
\EQ
A_0={\sqrt{\alpha'}\over g_{YM}^2}\, {U^7 \over Q} , 
\label{eq310}
\EN
respectively, where the charge $Q$ is now redefined as 
\EQ
Q=60\pi^3 (\alpha')^{-3/2} g_sN=240\pi^5 g_{YM}^2 N .
\EN
Since the solution is static, the spatial components of the 
RR gauge field are zero. 

The 11 dimensional metric in the near horizon limit 
 is invariant  under the 
same scale transformation as  before 
\EQ
U\rightarrow \lambda U ,
\label{eq312}
\EN
\EQ
t \rightarrow \lambda^{-1}t ,
\label{eq313}
\EN
\EQ
g_s \rightarrow \lambda^3 g_s ,
\label{eq314}
\EN
if we assume that the 11th coordinate $x_{11}$ is invariant. 

Furthermore, the 10 dimensional metric (\ref{eq38}) and 
the dilaton (\ref{eq39})  are  invariant under the 
special coordinate transformation whose infinitesimal form is 
\EQ
\delta_K t = -  (t^2 +k{g_{YM}^2\over U^5}) , 
\label{eq315}
\EN
\EQ
\delta_K U =2  tU ,
\label{eq316}
\EN
\EQ
\delta_K g_s=6  t g_s ,
\label{eq317}
\EN
where $k$ is a constant independent of the string coupling 
$k=96\pi^5 N$.  
Together with the trivial time translation $\delta_H t =1, 
\delta_H U =0, \delta_H g_s =0$ and the scaling 
$\delta_D t =-t, \delta_D U =U, \delta_D g_s = 3g_s$, 
we again have the SU(1,1) algebra (\ref{eq210}). 
The gauge field $A_0$ transforms as a conformal 
field of dimension 1. 

Since at the asymptotic boundary $U\rightarrow \infty$, the 
extra $U$ dependent part of the 
special conformal transformation 
(\ref{eq315}) vanishes, it is consistent to interpret the 
transformations (\ref{eq27})$\sim$(\ref{eq29}) 
 of the Yang-Mills quantum mechanics 
as the symmetry corresponding to the transformation 
at the boundary of the near horizon geometry described by the 
background fields (\ref{eq38})$\sim$(\ref{eq310}).  
The situation is similar to the 3+1 dimension Yang-Mills 
theory. 

It should be kept in mind that the 
space-time geometry of the 10 dimensional 
metric (\ref{eq36}) is not the AdS$_2$
$\times S^8$, since the would-be radius of the anti-de Sitter 
space $\propto (g_{YM}^2N/U^3)^{1/4}$ is not a constant.  
The above result, however, strongly indicates that the 
Yang-Mills matrix model may be 
interpreted as the boundary conformal field theory of the 
type IIA supergravity, in essentially the same sense 
as for the 3+1 dimensional Yang-Mills theory and the ADS$_5$$\times S^5$ space-time, provided we properly  
take into account the dilaton which is now space-time dependent. 
As discussed in \cite{maldacena2}, the correspondence is expected to be valid 
in the region 
\EQ
g_{YM}^{2/3}N^{1/7} \ll U \ll g_{YM}^{2/3} N^{1/3}
\label{condition1}
\EN
where the first and second inequalities come from 
the weak coupling condition $e^{\phi} \ll 1$ and the 
small curvature condition $R\sim (g_sN)^{-1/2}U^{3/2}(\alpha')^{-1/4} \ll (\alpha')^{-1}$, 
respectively. 
In terms of  the original coordinate $r$ and the string coupling, 
the condition is 
 \EQ
\sqrt{\alpha'} g_s^{1/3}N^{1/7} \ll r \ll \sqrt{\alpha'}(g_sN)^{1/3}. 
\label{condition2}
\EN
The near-horizon condition requires  
$\,  r\ll \sqrt{\alpha'}(g_sN)^{1/7}$.  For sufficiently large $N$ and 
small $g_s$ with large $g_sN$, there is an overlap region 
for the validity of these conditions.  Note that the 
conditions (\ref{condition1}) (or (\ref{condition2})) are dilatation invariant, 
while the near horizon condition is not. 
However, the latter is automatically satisfied in the 
Maldacena limit.

On the other hand, 
 it should be noted that the range for the validity of the 
naive loop expansion in the matrix model is 
$U > g_{YM}^{2/3} N^{1/3}$.  Thus, there seems to be
 no overlap with the region where we can trust the results of naive loop expansion. However, if we are only interested in the  ``classical" 
contribution expanded in the special combination 
${G_{11}N\over R_{11}^2}r^{-7}\bigl({dr\over dt}\bigr)^2$, 
where $G_{11}\propto g_s^3\ell_s^9, R_{11}=g_s\ell_s$ 
are the 11 dimensional Newton constant and 11 dimensional 
compactification radius, respectively, 
the loop expansion can be compared 
to the supergravity  in the limit of very 
small velocity \,   $v^2 \ll {(\alpha')^{7/2}r^7 \over g_sN}$. 

\vspace{0.5cm}
\section{The conformal symmetry and D-particle interactions} 
\setcounter{equation}{0}

We next show that the conformal symmetry found above 
puts a strong constraint on the effective action 
for D-particle interactions.  In the case of D3-brane, 
it was argued in \cite{maldacena} that the anti-de Sitter 
conformal symmetry, combined with a 
supersymmetric non-renormalization theorem, 
determines the bosonic part of the Born-Infeld action 
of a D3-brane in the AdS background. 
What we will establish in the following is 
the counterpart of this result for the case of D-particle. 

Let us consider the scattering of a probe D-particle 
in the background of the source system with a large number $N$ of 
coincident D-particles.  Let the distance between the 
probe and the source be $U(t)$ using the same 
convention as in the last section. 
If we neglect the possible acceleration 
dependent terms in the effective action and 
consider only the motion along the radial direction $U$
for simplicity, the 
time translation and the scaling symmetry (\ref{eq312})$\sim$
(\ref{eq314}) 
restricts the form of the bosonic effective action 
into 
\EQ
S_{eff} =\int \, dt\, {1\over 2g_s}\bigl({d U\over dt}\bigr)^2F\Bigl({g_s \over U^3}, {1\over U^4}
\bigl({d U\over dt}\bigr)^2\Bigr)
\EN
where we have assumed invariance under time inversion too. 
This form has been already known from the work  \cite{bbpt}. 
Now under the special conformal transformation,  
the field $U(t)$ and the string coupling $g_s$ are transformed as 
\EQ
U(t) \rightarrow U'(t) =(1+2\epsilon t)U(t') , \label{eq42}
\EN
\EQ
t'=t+\epsilon(t^2 + {a\over U^5}) , \label{eq43}
\EN
\EQ
g_s \rightarrow (1+6\epsilon t)g_s , \label{eq44}
\EN
to the first order with respect to the infinitesimal parameter 
$\epsilon$. Note that the sign in (\ref{eq43}) which is 
due to our use of $t$, instead of $t'$,  in $U'(t)$ and 
$dt'=(1+2\epsilon t -\epsilon{5a\over U^6}{dU\over dt} )dt$. 
Here we abbreviated as $a\equiv kg_{YM}^2
=24\pi^3 g_s (\alpha')^{-3/2} N$.  The transformation law 
of the velocity  is thus 
\EQ
{dU(t) \over dt} \rightarrow (1+4\epsilon t) {dU(t')\over dt'} 
+2\epsilon U(t') - \epsilon {5a \over U(t')^6}
\bigl( {dU(t')\over dt'}\bigr)^2 , 
\label{eq45}
\EN
\EQ
\bigl({dU(t) \over dt}\bigr)^2 
\rightarrow 
(1+8\epsilon t) \bigl({dU(t')\over dt'}\bigr)^2
+4\epsilon U(t'){dU(t')\over dt'} - \epsilon {10a \over U(t')^6}
\bigl( {dU(t')\over dt'}\bigr)^3 ,
\label{eq46}
\EN
\[ .... etc. \] 
This shows that while the combination $g_s/U^3$ 
is invariant under the special conformal 
transformation, ${1\over U^4}
\bigl({d U\over dt}\bigr)^2$ is not invariant. 
In particular,  if we expand the function $F$ 
in the double expansion with respect to 
these two combinations, the velocity expansion 
must appear as a power series with respect to the special combination 
\EQ
 w\equiv {g_s\over U^7}
\bigl({d U\over dt}\bigr)^2
\EN
and then the coefficients of the expansion 
are uniquely determined by the 
coefficient of the first term. The reason for this 
is that , by the transformation rule (\ref{eq45}), 
$w^n$ generates 
$w^{n-1}U (dU/dt) \sim U^{-7(n-1) +1}(dU/dt)^{2(n-1) +1}$
and $w^{n-1}U^{-6}(dU/dt)^3 \sim U^{-7(n-1)-6}\\
\times (dU/dt)^{2(n-1)+3}$.  This shows that the power of 
the squared velocity can increase only with the 
power of $U^{-7}$, and further allows us to fix the coefficients 
recursively from lower to higher $n$ up to an arbitrary 
function of $g_s/U^3$ which is common to all the 
coefficients.  

Thus, the effective lagrangian must 
take the form
\EQ
L= {1\over 2g_s}\bigl({d U\over dt}\bigr)^2
f({g_s \over U^3})\sum_{n=0}^{\infty} c_n w^n .
\EN 
Note that we automatically get a factorized form. 
However, the supersymmetric non-renormalization theorem 
 for supersymmetric particle mechanics 
\footnote{
For a recent discussion on the constraints from the 
super symmetry, see \cite{paban} where the often stated 
folklore theorem about the $v^2$ term is explicitly proven. 
}tells us that the coefficient of the first term $\bigl({d U\over dt}\bigr)^2$  should be the same 
as the classical one, namely a constant independent of $U$. 
We can thus set $f=$constant.  The first few coefficients 
determined by applying the transformation (\ref{eq45}) are 
\EQ
c_0=1 , \, \, \, 
c_1={15\over 16}N({\alpha'\over 4\pi^2})^{-3/2} , \, \, \, 
c_2 = {225\over 64}N^2 ({\alpha'\over 4\pi^2})^{-3} , \,  \, \, \ldots .
\EN
In this form, as discussed in the end of the 
last section, the result can be compared with the loop 
expansion at least for sufficiently small 
velocity and it is known that the results agree up to 2 loops 
\cite{bb}.  
The effective action can be rewritten 
in the closed form which was derived in \cite{bbpt}  
\EQ
S_{eff}= -\int\, dt\, {1\over R}{\sqrt{1-h_{--}(r)\bigl({dr\over dt}\bigl)^2} -1 \over 
h_{--}(r)}
\label{eq49}
\EN
with 
\EQ
h_{--}(r) = {Q(\alpha')^5 \over r^7}
\EN
where we have returned to the original coordinate $r$ instead 
of the energy $U$ of open strings. By comparing with the 
convention of ref.\cite{bbpt}, 
$h_{--}(r) = {15N\over 2R^2M^9 r^7}$ where $R=g_s \ell_s$ and 
$M^{-1}=g_s^{1/3}\ell_s$ are the compactification radius along the 
light-like 11th direction and the 11 dimensional Planck mass, respectively, 
we have $\ell_s=(2\pi)^{3/7}\sqrt{\alpha'}$ in our convention.  
It is easy to directly check that 
the effective action is indeed invariant up to a total 
derivative under the special 
conformal transformation (\ref{eq42})$\sim$(\ref{eq44}). 
We also note that although we have only considered the 
motion along the radial direction for simplicity, the final 
form of the effective action can be extended to 
the general case by making the replacement 
$(dr/dt)^2 \rightarrow v^2=(dx_i/dt)^2$ because of rotation invariance 
in the transverse space.

The above result is not trivial, since the 11 dimensional metric itself 
is not invariant under the conformal transformation. 
Only after taking the near horizon limit for the 
10 dimensional metric, we have the conformal 
invariance.  On the other hand, in the original derivation \cite{bbpt} 
of the above effective action,  the choice of the light-cone frame was essential.  
We interpret this result 
as an indication  that 
the Maldacena limit for D-particles in 10 dimensions has the 
same effect with respect to the conformal symmetry as caused by 
going to the light-cone frame for the massless 
plane wave in 11 dimensions.  Remember that, 
if the compactification were performed in the space-like 
11th direction instead of the light-like direction,  
we would have obtained a different action in which $h_{--}$ is 
replace by $1+h_{--}$.  This may be regarded as
evidence  for the fact 
that the Maldacena limit and the discrete light-cone 
prescription \cite{suss} have a common range of validity and can be 
smoothly connected to each other. 
This also suggests that the conformal symmetry is present 
even in the region $U < g_{{\rm YM}}^{2/3}N^{1/9}$ of Matrix black holes as discussed in 
\cite{maldacena2} where the plane-wave description 
of the 11D classical solution can no more be trusted. 

\vspace{0.5cm}
\section{Concluding remarks}

First we note that 
our results partially explain, as a consequence of the 
conformal symmetry,  
why the BFSS matrix model gives the  
classical effective action of the same form as 
derived from the supergravity theory in 11 dimensions. 
However, it should be 
emphasized that our discussion does not 
prove that we must have the complete agreement to all orders,  
since we have not shown explicitly that 
the field dependent transformation of the form 
(\ref{eq315}) is realized for the probe D-particle 
in the matrix model.   It is possible that, in the matrix model, 
the field dependent transformation for the probe D-particle 
might become more complicated due to the $\alpha'$ 
and quantum-gravity corrections,  
if the Matrix theory conjecture is correct, and 
hence the form of the effective action might be subject to further 
corrections. 

It should in principle be possible to derive the field dependent 
special conformal transformation from the linear 
transformation law (\ref{eq29}).  For example, 
we can separate the operator $\tilde{K}$ into 
the diagonal and off-diagonal contributions. 
\EQ
\tilde{K} = \tilde{K}_{diagonal} + \tilde{K}_{off-diagonal}
\EN
where 
\EQ
 \tilde{K}_{off-diagonal}= -{1\over g_s}\sum_{a\ne b}|X_{ab}|^2 .
\EN
The one-loop expectation value of this quantity 
in the background with a large number of coincident 
source D-particles and a probe D-particle at the distance $U$ is 
expanded in the form 
\EQ
\langle \sum_{a\ne b}|X_{ab}|^2\rangle 
\propto \sum_n c_n {g_s\over U}\Bigl({1\over U^4}\bigl({dU\over dt}\bigr)^2\Bigr)^n. 
\EN
We can see that the 
$n=1$ term is just consistent with the field dependent 
transformation of the form (\ref{eq43}) or (\ref{eq45}).  
The question is then whether or not the whole off-diagonal 
contributions can be represented by this term 
$U^{-5}\bigl({dU\over dt}\bigr)^2$ in the operator formalism. 
We have no clear answer at present.  We emphasize that 
there are similar questions 
for the D3-brane, but answer is not known.  
It would be extremely interesting if there is a systematic 
way of determining the field dependent symmetry 
transformations for arbitrary background from the Yang-Mills 
matrix models alone.  It would 
amount to the proof of the Maldacena conjecture 
and would be of great help for investigating 
the dynamics of D-branes in the matrix models 
beyond the usual loop computations in the semi-classical 
approximation. 

Finally, we mention other remaining problems:
\begin{enumerate}
\item Extension to other D$p$-branes ($p\ne 0, 3$): From the discussions in section 2, it should be more or less clear that the idea  
behind the conformal symmetry, namely 
the space-time uncertainty principle,  is rather general and the 
discussions of the present paper are not restricted to D-particles. 
Most of the results can be extended to other cases without much difficulty, except for the problem of non-renormalizable 
(perturbatively, at least) 
world-volume theories for $p\ge 4$. 
\item  Extension to superconformal symmetry: In view of  the result 
of the last section, we may hope that the 
supersymmetric extension of the conformal symmetry 
will put very powerful constraints on the D-particle dynamics. 
In particular, we may have generalized non-renormalization 
theorems by using the super-conformal Ward identities. 
\item  Geometrical interpretation of the conformal transformation: 
Although the near-horizon geometry is not AdS,  our results 
suggest that there might be some geometrical 
structure if one properly takes into account the nontrivial 
behavior of the dilaton and string coupling constant. 
\item  Comparison of Green functions and spectrum
 between  the supergravity and 
matrix model:  Since the dilaton is not constant, 
computation will be more complicated than the case 
of D3-brane and other non-dilatonic branes
 \cite{klebanov0, klebanov, witten}.   
Such study is expected to be useful for 
understanding the non-perturbative 
structure of the large $N$ dynamics of the Matrix theory. 
\item Dynamics of the dilaton and the string coupling: 
The time translation generator of our conformal 
transformations includes a kinetic term for the 
string coupling. Therefore, if we adopt it as the 
hamiltonian for time evolution of the system, there is a 
nontrivial dynamics for the ground-state value 
of the dilaton which is 
coupled to the dynamics of D-particles. 
It is interesting to see whether or not the extended hamiltonian 
really captures the dynamics of the dilaton. 
For example, the dilaton kinetic term might be 
related to the zero mode of light-cone formulation. 
\item Elimination of string coupling constant: Another question 
related to the dilaton is whether it is possible to 
eliminate the string-coupling constant from the 
formalism. For example, in the string field theory formalism, 
we can indeed eliminate the string coupling 
from the action by making a shift for the string field as 
discussed in \cite{ty4}. If we can reformulate the matrix 
theory in a similar way, we would not have to perform the conformal 
transformation for the string coupling explicitly, and would 
be able to give a more satisfactory formulation 
of the conformal symmetry as a realization of the 
space-time uncertainty principle.  
This question is closely related to the problem of background 
(in)dependence of the Yang-Mills matrix theory. 
\end{enumerate}
Some of these issues will be discussed in a forthcoming 
paper \cite{jky}. 

\vspace{1cm}
\noindent
{\bf Acknowledgements}

The main part of the present work was done during our stay   
at ITP, Santa Barbara, participating in the 
workshop, ``Dualities in String Theory". 
We would like to thank ITP 
for its hospitality and  the members of the 
workshop for stimulating discussions and, especially, 
 Y. Kazama for  his useful comments 
 on  the present work and the manuscript.   
The research at ITP was supported in part by the National 
Science Foundation under Grant No. PHY94-07194. 
The work of A. J. is supported in part by  the Department of Energy under contract DE-FG02-91ER40688-Task A.
The work of T.Y. is supported in part 
by Grant-in-Aid for Scientific  Research (No. 09640337) 
and Grant-in-Aid for International Scientific Research 
(Joint Research, No. 10044061) from the Ministry of  Education, Science and Culture.

\end{document}